\documentclass[prd,aps,floats,twocolumn,nofootinbib]{revtex4-1}

\usepackage{graphicx}
\usepackage{epsfig}
\usepackage{latexsym}
\usepackage{amssymb}
\usepackage{bm}
\usepackage{float}
\usepackage{subfigure}
\usepackage{amsmath}
\usepackage{amsfonts}
\usepackage{bm}

\newcommand{\be}{\begin{equation}}
\newcommand{\ee}{\end{equation}}
\newcommand{\bea}{\begin{eqnarray}}
\newcommand{\eea}{\end{eqnarray}}

\usepackage[normalem]{ulem}

\begin{document}

\title{
Cosmological time and the constants of nature
}

\date{\today}

\newcommand{\addressImperial}{Theoretical Physics Group, Imperial College, Prince Consort Rd, London, SW7 2BZ, United Kingdom}

\author{Jo\~{a}o Magueijo}\affiliation{\addressImperial}
\email{j.magueijo@imperial.ac.uk}

\begin{abstract}

We propose that cosmological time is {\it effectively} the conjugate of the constants of nature.
Different definitions of time arise, with the most relevant related to the constant controlling the dynamics in each epoch. The Hamiltonian constraint then 
becomes a Schrodinger equation. In the connection representation, it is solved by monochromatic 
plane waves moving in a space generalizing the Chern-Simons functional.
Normalizable superpositions exist and for factorizable coherent states we recover the classical limit 
and a seamless handover between potentially disparate times. There is also a rich structure of alternative states, 
including entangled constants, opening up the doors to new phenomenology.
\end{abstract}

\maketitle

The seemingly unrelated problems of time in Quantum Gravity~\cite{Isham,Kuch} and of the values of the constants 
of nature~\cite{Barrowtip} are two foundational nuisances in physics occasionally dismissed as metaphysical. 
And yet, without a physical definition of time it is hard to see how Quantum Gravity could ever connect with the 
real world. Perhaps for this reason, the subject is often relegated to the depths of the ``Planck epoch'', as if our classical Universe
should not appear as a limit of Quantum Cosmology, possibly with perennial corrections and novelties, and so direct observational consequences. The fact that we appear to be entering a baffling phase of accelerated expansion~\cite{accelexp}, possibly due to the cosmological ``constant''  $\Lambda$ (with all its fine-tuning headaches~\cite{weinberg,padilla})  might be taken as a hint. Could the two puzzles be closely related?



In this Letter we propose adopting as effective cosmological clocks 
the conjugates of the constants of nature.
 For this to be possible we have to partly ``deconstantize'' them, demoting constants from parameters 
``set in stone'' 
to phase space variables which happen to be
constant as a result of the equations of motion (thus, leaving room for an explanation of their values).  
This proposal is intuitive: it is only natural that the conjugate (and quantum complementary) of something that does not vary---a constant---should be the essence of change itself, and so a measure of relational, physical time (as opposed to coordinate time). It is also not entirely new: in the context of $\Lambda$ it  resonates with unimodular gravity~\cite{unimod1,unimod}, where the conjugate of $\Lambda$ was identified with a time variable (which turned out to be Misner's volume time~\cite{misner}). 

But why would $\Lambda$ be special? A clock is crafted with what is at hand.  Depending on the constant(s) dominating the dynamics
in each region of phase space, we should employ different times. We are thus led to {\it emergent} times variously conjugate to 
$\Lambda$,  the gravitational constant, $G_N$, or even
the speed of light, $c$, among others. 
This is usually regarded as a curse: either there is no time in General Relativity, or, if we attempt to define one, we are spoilt for choice. In this Letter we 
embrace a democracy of times as a blessing, with the adjustment of clocks across different ``time zones'' 
to be seen as a physical  feature of our world.
Semiclassical states will be found, for which the handover between clocks is trivial. But there are also alternative states, which might be 
of great predictive relevance, considering that the Universe is currently busy passing on the baton from a matter (or $G_N$) clock to a $\Lambda$ clock.

We focus on constants, $\alpha$, for which the timeless Hamiltonian constraint $H = 0$ 
can be  {\it approximately} written as:
\be\label{splitH}
{\cal H}= H_0 - \alpha = 0
\ee
in some phase space regions. 
We illustrate our approach in minisuperspace, but generalizations 
can be built~\cite{gens}, using unimodular gravity as the blueprint~\cite{unimod,gielen} (but also 
non-local theories~\cite{pad,lomb}).
We use the Einstein-Cartan formulation of Relativity, but 
our construction can be applied to other theories of gravity.  Then our starting point is the action:
\begin{equation}\label{S}
S=\frac{3 V_{c}}{8\pi G_{0}} \int dt\bigg(a^{2}\dot{b}-Na \left[- (b^{2}+k c^2
)+ \sum_i\frac{m_i}{a^{1+3w_i}}\right]\bigg),
\end{equation}
where $G_{0}$ is a fixed value for $G_N$ (should $G_N$ be up for ``deconstantization''), 
$a$ is the expansion factor, $b$ is the connection variable ($b= \dot a/N$ on-shell), 
$k=0,\pm1$ is the normalized spatial curvature, $N$ is the lapse function and $V_c$ is the comoving 
volume of the region under study (or the whole manifold, if compact). 
The summation  term 
can accommodate a large number of models, the specifics of which 
will not be relevant for the rest of this Letter, but which we list here nonetheless. 
For a geometrical $\Lambda$ we have 
$m_i=\Lambda/3$ and  equation of state $w_i=-1$. For dust and radiation we have 
$w_i= 0,1/3$, and can set $m_i=C_i 8\pi G_0/3$, where $C_i$ are conserved quantities~\cite{gielen}.
We can also set  $m_i=C_i 8\pi G_N/3$, and deconstantize the gravitational coupling $G_N$
(as opposed to the $G_0$ controlling the non-commutation properties of 
geometrical operators). For a massless scalar field, in particular, we would then have $m_i\propto \Pi_\phi/G_N$
and $w_i=1$. 
Finally, and more speculatively, both the curvature $k$ and $\Lambda$ could be seen as 
components with $m_i= - k c^2 $ and $w_i=-1/3$, and $m_i=\Lambda_0 c^2/3$ and $w_i=-1$, with $c$ the target constant. 

In this Letter we develop the quantum theory in the connection representation (but metric 
versions exist~\cite{gielen}, with the two equivalent if it is accepted that $a^2\in(-\infty,\infty)$~\cite{CSHHV}). Then, the Hamiltonian constraint
can be more easily solved by putting (\ref{splitH}) in the approximate form:
\be\label{splitH1}
{\cal H}= h_i(b)a^2 - \alpha_i = 0,
\ee
in each region/epoch  where one $i$ in (\ref{S}) dominates (we will examine each $i$ separately at first;
no summation over $i$ implied where repeated).
As we shall see, this amounts to a canonical transformation selecting 
``linearizing'' variables (to use the terminology of~\cite{DSR}).
From (\ref{S}) we see that:
\bea
h_i(b)&=&(b^2+k)^{\frac{2}{1+3w_i}}\\
\alpha_i &=& m_i^{\frac{2}{1+3w_i}},
\eea
(with $w_i=-1/3$ to be treated separately elsewhere). 
The central proposal in this Letter
is to promote the $\alpha_i$ to phase space variables
with conjugate momenta $p_i$ to be identified as a time variable $T_i\equiv p_i$:
\be\label{Sextension}
S\rightarrow S+
\frac{3 V_{c}}{8\pi G_{0}} 
\int dt \dot\alpha_i p_i,
\ee
(the pre-factor introduced for later convenience). In this perspective, 
the constancy of the ``constants'' is not preset,  but rather arises from the
equation of motion:
\be
\dot \alpha_i=\{\alpha_i,H\}=0
\ee
resulting from Poisson bracket:
\be\label{PB}
\{\alpha_i,p_i\}=\frac{8\pi G_{0}}{3V_c}
\ee
and the fact that  $H$ is independent of $p_i$ (in itself an expression of the ``time''-independence of the 
Hamiltonian, at a more fundamental level). Classically,  
the $\alpha_i$ values are just integration constants, generalizing a  well-known situation in unimodular gravity~\cite{unimod1,unimod} and not
only~\cite{Hawking3f}. But more importantly, given~(\ref{PB}), 
the quantum theory now reveals an {\it effective} Schrodinger  equation
\be\label{WDWSchro1}
\left[H_0(b)- i \frac{l_P^2}{3V_c} \frac{\partial }{\partial T_i}\right]\psi(b,T_i) =0
\ee
(with fixed $l_{P}=\sqrt{8\pi G_{0}\hbar }$)
in lieu of the timeless Wheeler-DeWitt (WDW) equation.
It is only when we impose ``fixed constants'', 
i.e. the monochromatic ansatz:
\be\label{monoch}
\psi (b,T_i;\alpha_i)= \psi_s(b;\alpha_i ) \exp{\left[- i \frac{3V_c}{l_P^2}\alpha_i T_i \right]},
\ee
that we find that the  ``spatial'' $\psi_s$ must satisfy 
the original timeless WDW equation.  Furthermore, since:
\be\label{a2inb}
\{b,a^2\}=\frac{8\pi G_{0}}{3V_c}\implies 
\hat a^2=-i\frac{l_P^2}{3V_c}\frac{\partial}{\partial b}
\ee
we see that (with
$H_0\approx h_i(b)a^2$) the WDW equation, with suitable ordering, becomes:
\be
\left(-i\frac{l_P^2}{3V_c} \frac{\partial}{\partial X_i}-
 \alpha_i
\right)\psi_s =0.
\ee
with:
 \be\label{X}
X_i (b)=\int \frac{db}{h_i(b)}.
\ee
The ``spatial'' solutions are therefore plane-waves in $X_i$:
\bea
\psi_s(b;\alpha_i)&=& {\cal A}(\alpha_i) \exp{\left[i\frac{3V_c }{l_P^2} \alpha_i X_i(b)   \right]}\label{psi0}
\eea
where for $w_i=-1$ we can recognize the Chern-Simons state~\cite{CS,jackiw,kodama}. In this sense 
Eq.~(\ref{X}) is a generalization of the Chern-Simons functional. The full monochromatic solutions are:
\bea
\psi (b,T_i;\alpha_i)&=& {\cal A}(\alpha_i) \exp{\left[i\frac{3V_c }{l_P^2} \alpha_i (X_i(b) - T_i)  \right]}.\label{planew0}
\eea
These are not plane-waves in $b$ or in the original $m_i$, indeed in terms of them the waves
have non-linear dispersion. The $X_i$ and $\alpha_i$ are ``linearizing'' variables~\cite{DSR,DSR1},
in terms of which the solutions become plane waves moving with constant speed in superspace. 

By demoting the $\alpha_i$ to circumstantial constants we gain more than a time variable in the quantum theory:
we enlarge the space of solutions. Instead of being restricted to 
(\ref{planew0}) 
we are allowed superpositions:
\be\label{wavepackets}
\psi (b,T_i)=\int d\alpha_i {\cal A}(\alpha_i) \exp{\left[i\frac{3V_c }{l_P^2}\alpha_i (X_i(b) - T_i)  \right]}.
\ee
Among these, we highlight coherent/squeezed states, with
${\cal A}(\alpha_i)=\sqrt{{\bf N}(\alpha_{i0},\sigma_i)}$ (where $\bf N$ denotes a normal distribution)
resulting in:
\bea\label{coherent}
\psi_S(b,T_i)&=&
%
{\cal N} \psi(b,T_i;\alpha_{i0})\exp\left[-\frac{\sigma_i^2(X_i-T_i)^2}
{(l_P^2/3V_c)^2}
\right]
\eea
where $\sigma_i$ can be dialled anywhere between the limiting cases of a plane wave 
(${\cal A}\propto \delta(\alpha-\alpha_{i0})$)
and the ``light-ray'' $\psi\propto \delta(T_i-X_i$) following from ${\cal A}\propto {\rm const}$
(which is actually required in quasi-topological theories~\cite{GB,MZ}, and expresses the conformal constraint).
Note the Heisenberg uncertainty relation between each $\alpha$ and its associated time:
\be\label{heis}
\sigma_{T}\sigma_{\alpha}\ge \frac{ l_P^2}{ 6V_c }.
\ee
Coherent states $\psi_C(b,T_i)$ are states (\ref{coherent}) with:
\be
\sigma^2_i = \frac{ l_P^2}{6V_c},
\ee
saturating (\ref{heis}) and equally spreading the uncertainties.

In fact, 
we do not even need to construct general 
states from monochromatic waves. 
Eq.~(\ref{WDWSchro1}) can be rewritten as:
\be
\left(\frac{\partial}{\partial X_i}-
 \frac{\partial}{\partial T_i}
\right)\psi =0,
\ee
(suggesting the conserved current, 
$
j^0=j^1=|\psi|^2
$, 
with the concomitant probability interpretation). Then, the 
general solution is of the form:
\be\label{FofTmX}
\psi(b)=F(T_i -X_i).
\ee
By setting $F$, for example, a Gaussian, or any other such localized function, we thus have an expression
for a solitonic wave, showing no dispersion in $X$. All of these solutions are normalizable with the ``naive'' inner product.
No longer do we need to blame the trivial inner product
for the non-normalizability of the monochromatic solutions.

Having a time variable and this larger space of solutions 
is of paramount importance for making contact with reality. 
Plane waves imply a uniform distribution in $b$, hardly a prediction, but they are also 
not immediately physical.   
We need {\it both} a time variable {\it and} 
the ability to superpose plane waves into normalized peaked distributions to recover something minimally physical.
In fact the semi-classical limit is recovered for the coherent states $\psi_C(b,T_i)$, for which the 
second Hamilton equation:
\be\label{secondH}
\dot T_i =\{T_i,H\}=-\frac{1+3w_i}{2}Na^{-3w_i} \alpha_i ^\frac{3w_i-1}{2}
\ee
is true not only on average (an expression of Ehrenfest's theorem), but with minimal and balanced 
uncertainties in the complementary $\alpha_i$ and $T_i$ appearing on the two sides of (\ref{secondH}).
Their peak, then, follows the classical trajectory since it is easy to check that 
$\dot T_i=\dot X_i$, with (\ref{secondH}) and (\ref{X}), 
is equivalent to the Einstein equations\footnote{Note the difference of interpretation with regards to 
Chern-Simons time~\cite{Chopin} for $w_i=-1$. Here $X_i=\Im(Y_{CS})$ is not a time, but a spatial variable. Time, instead, is the conjugate of $\alpha=3/\Lambda$. The two can be loosely confused only 
because  if $\psi$ is peaked,  its peak moves along the outgoing ``light-ray'' $T_i=X_i$.}.

But now we have an embarrassment of riches.
Let $\bm{\alpha}$ be a vector representing the whole set of 
relevant constants and  $\bm{T}$ their conjugates. The various components of $\bm{T}$
are a priori independent variables, so we have a plethora of times instead of a single one. 
True, (\ref{secondH}) implies that  classically and on-shell they are all related by a lapse function redefinition.
Indeed, we see that for  $\Lambda$, radiation and dust, $T_i$ is proportional to volume time~\cite{misner}, 
conformal time and cosmological proper time, respectively.  Quantum mechanically, however,
the situation is more complex.

The implication in this Letter is that 
there is not one ``Schrodinger'' equation (our examples being 
merely limiting cases) but a PDE in multiple times running concurrently,
obtained by taking the Hamiltonian in  (\ref{S}) and applying the replacements:
\be
H\left[b,a^2;\bm \alpha \rightarrow  i \frac{l_P^2}{3V_c} \frac{\partial }{\partial \bm T}\right]\psi=0.
\ee
Its general solutions are:
\be\label{gensol}
\psi(b)=\int d\bm{\alpha} {\cal A}(\bm \alpha) \exp{\left[-i\frac{3V_c }{l_P^2}\bm \alpha \bm T  \right]}\psi_s(b;\bm \alpha),
\ee
where $\psi_s(b;\bm \alpha)$ solves the WDW equation with constant $\bm\alpha$. 
However, for a Hamiltonian carving up phase space into regions dominated by a single constant, the readjustment of quantum clocks 
across such regions is seamless if we assume coherent states in all $\alpha_i$ {\it and}  factorization:
\be\label{factcoh}
{\cal A}(\bm \alpha) =\prod_i\sqrt {{\bf N}(\alpha_{0i},l_P^2/6V_c)}.
\ee
Then, 
the $\psi_s(b;\bm \alpha)$ is a piecewise plane wave in the $X_{i}(b)$ associated with each dominant $\alpha_i$. 
Each piece engages with the  phase associated with the corresponding $T_{i}$ 
(hence, the approximate single-time Schrodinger equation),
producing a wave-packet describing the correct classical limit, as above.  If ${\cal A}(\bm \alpha)$ factorizes, 
all the other times factorize too, and stop describing the $X$ evolution.  We end up with a classical 
limit in terms of different $X$ and $T$ in each region, but classically they are all equivalent. 

The semiclassical limit is reassuring, but our proposal also permits a wild west of alternative  states, wherein lies its true novelty. For example, we could set up a state made up of the superposition of coherent packets 
centred on different ${\bm \alpha}_0$  and shifted by various ${\bm T}_0$:
\be\label{ensemble}
\psi_E=\sum_{{\bm\alpha}_0,{\bm T}_0} \psi_C(b,{\bm T}- {\bm T}_0;{\bm \alpha}_0).
\ee
Such a {\it single} state provides the perfect realization of
an ``ensemble'' of Universes, with all combinations of constants realized, 
and even different  realizations (shifted in $\bm T$) for the same set of constants. Could this provide a solid basis for an explanation of the values of the constants~\cite{Barrowtip}?

But we could also consider non-semiclassical states. The possibility that ${\cal A}(\bm \alpha)$ does not factorize,
so that the constants are in an entangled state, adds a layer of complexity to such an ensemble.  Moreover
entangled states would leave a memory of non-dominant times at any epoch, since they would not factor out 
in (\ref{gensol}), as they do in the argument above (with phenomenology currently under investigation~\cite{future}). 
We could also consider hybrid states, for example a factorizable state which is
coherent in $G_N$ (associated with radiation and matter domination), but squeezed or even plane-wave-like in $\Lambda$
(and so ``smudged'' in $\Lambda$-time).
Given the prospect of imminent $\Lambda$-domination~\cite{accelexp}, this should raise an eschatological alarm. 
Could Quantum Cosmology be in the future, and the future be timeless~\cite{future}?

Even avoiding such extremes, new phenomenology may be inevitable. Note that 
the fact that the general solution (\ref{FofTmX}) 
contains outgoing-only waves is a reflection of the ``horizon problem'' (recall that on-shell $b= \dot a/N$, i.e. 
the inverse of the comoving Hubble scale). Indeed $\dot T=\dot X$ 
implies that a positive $b$ must decrease if $w>-1/3$, and increase otherwise. We therefore have a ``bounce''
in connection space each time the dominant equation of state crosses $w=-1/3$. 
Thus, 
from the point of view of our proposal, we are living in ``interesting times'' for two reasons. 
First, components with different $w$ currently have comparable densities, implying that the Universe is 
in the process of switching clocks (from a dust/$G_N$ clock to a $\Lambda$ one). Second,
the change in $w$ is such that we must have just emerged from a bounce in $b$, something that can only have 
happened on another occasion 
if there was a period of primordial inflation.

The full implications will be studied elsewhere~\cite{future}, 
but here we illustrate the issues with
a toy model involving $\Lambda$ and radiation. Setting $g=b^2+k$, we have $g^2\ge g_0^2=\frac{4}{3}\Lambda m_r$, and with $\alpha=3/\Lambda$ in (\ref{splitH}) we find after some algebra plane waves in:
\be
X(b;\Lambda m_r,\pm)=\int db\frac{1}{2}\left(g\pm\sqrt{{g^2-\frac{4}{3}\Lambda m_r}}\right)
\ee
where the $+/-$ refer to $\Lambda$/radiation domination.  It is not hard to then reproduce the argument 
presented above for a seamless handover of times when $g\gg g_0$ for the two branches. However, for $g\approx g_0$, even with 
(\ref{factcoh}), we see that (\ref{gensol}) does not factorize (due to the $X$ dependence on $\Lambda m_r$), with possible observational implications. 

Other, more speculative novelties may exist. Even in regions where one
ingredient $i$ dominates, there may be more than one constant affecting the dynamics. By construction, 
or by judicious choice of the state, it may be that only one of the associated times is relevant. But 
there are also more complex states juggling several times in the same region of phase space. 
This is particularly relevant for scalar fields in models with deconstantized $G_N$.
Another possible peculiarity involves models with bounces in $b$, even well away from the bounce.  
These models will see the wave packet pass a given $b$ more than once.
Classically this happens at different  ``times'', but quantum mechanically we have the inevitable interference terms. 
Could there be cross-talk between different times due to them? 
Even more speculatively, what would the situation be in an ensemble
of the form (\ref{ensemble}), where a given $b$ might be crossed by different ``realizations'' moving in opposite 
directions?

We close by briefly commenting on work akin to ours. We stress that our proposal is related, but not a twin of the 
suggestion that some constants may be complementary to each other~\cite{LeeGL,eigenconstant}. Here,
the complementary of a constant is, rather, a hypothetical momentum, which turns out to be a measure of time.
Our proposal is also orthogonal to the idea of ebb and flow in Chern-Simons time~\cite{ebb}, 
since here we interpreted the latter as  a spatial variable. Our formulation of unimodular gravity and its
generalization to other constants is also related to, but different from the ideas in~\cite{vikman} (a full formulation,
leading to the spatial constancy of $\alpha_i$ beyond minisuperspace is deferred to~\cite{gens}).
On the other hand we may make direct contact with the suggestion that the laws of physics could be evolving
(e.g.~\cite{selftaught}) and that time could be dynamical (e.g.~\cite{marina}, or even~\cite{rovelli}).
So as to introduce time variability in the laws of physics we must know what time is. Our definition, therefore, has immediate
implications upon this important discussion at the core of physics.


{\it Acknowledgments.} I would like to thank Stephon Alexander, Steffen Gielen, Chris Isham, Lee Smolin and Tom Zlosnik for discussions and advice. This work was supported by the STFC Consolidated Grant ST/L00044X/1.


\begin{thebibliography}{99}

\bibitem{Isham}
C.~J.~Isham,
``Canonical quantum gravity and the problem of time,''
NATO Sci. Ser. C \textbf{409}, 157-287 (1993).
\bibitem{Kuch} K. Kuchar,   ``Time and interpretations of quantum gravity," in 
Proceedings of
the 4th Canadian Conference on General Relativity and Relativistic Astrophysics',
World Scientic, Singapore, 1992.

\bibitem{Barrowtip} 
J.~D.~Barrow and F.~J.~Tipler,
``The Anthropic Cosmological Principle,'' Oxford University Press, 1988. 

\bibitem{accelexp} P.~J.~E.~Peebles and B.~Ratra,
Rev. Mod. Phys. \textbf{75}, 559-606 (2003).

\bibitem{weinberg}
S.~Weinberg,
Rev. Mod. Phys. \textbf{61}, 1-23 (1989).

\bibitem{padilla}
A.~Padilla,
``Lectures on the Cosmological Constant Problem,''
[arXiv:1502.05296 [hep-th]].

\bibitem{gens} C. Isham and J. Magueijo, in preparation. 

\bibitem{unimod1}
W. G. Unruh, Phys. Rev. {\bf D40}, 1048 (1989). 

\bibitem{unimod} M. Henneaux and C. Teitelboim, Physics Letters {\bf B 222}, 195
(1989).

\bibitem{misner}
C.~W.~Misner, Phys. Rev. {\bf 186}, 1328 (1969); Phys. Rev.1 {\bf 86}, 1319 (1969). 



%




\bibitem{gielen} 
S.~Gielen and L.~Men\'endez-Pidal,
Class. Quant. Grav. \textbf{37}, no.20, 205018 (2020).

\bibitem{pad}
N.~Kaloper and A.~Padilla,
Phys. Rev. Lett. \textbf{112}, no.9, 091304 (2014).



\bibitem{lomb}
L.~Lombriser,
Phys. Lett. B \textbf{797}, 134804 (2019).





\bibitem{CSHHV}
J.~Magueijo,
Phys. Rev. D \textbf{102}, no.4, 044034 (2020).


\bibitem{DSR}G.~Amelino-Camelia,
Int. J. Mod. Phys. D \textbf{11}, 35-60 (2002); 
J.~Magueijo and L.~Smolin,
Phys. Rev. Lett. \textbf{88}, 190403 (2002).

\bibitem{Hawking3f}
S.W. Hawking, 
Physics Letters B, {\bf 134}, 403 (1984). 

\bibitem{DSR1}
D.~Kimberly, J.~Magueijo and J.~Medeiros,
Phys. Rev. D \textbf{70}, 084007 (2004).

\bibitem{GB}
S.~Alexander, M.~Cort\^es, A.~R.~Liddle, J.~Magueijo, R.~Sims and L.~Smolin,
Phys. Rev. D \textbf{100}, no.8, 083506 (2019);
Phys. Rev. D \textbf{100}, no.8, 083507 (2019). 

\bibitem{MZ}
J.~Magueijo and T.~Z\l{}o\'snik,
Phys. Rev. D \textbf{100}, no.8, 084036 (2019).



\bibitem{CS}S. S. Chern and J. Simons, Ann. Math.  {\bf 99}, 48 
(1974);
G.~V.~Dunne,
``Aspects of Chern-Simons theory,''
[arXiv:hep-th/9902115 [hep-th]].

\bibitem{jackiw}
R.~Jackiw,
Conf. Proc. C \textbf{8306271}, 221-331 (1983)
MIT-CTP-1108.

\bibitem{kodama}
H.~Kodama,
Phys. Rev. D \textbf{42}, 2548-2565 (1990);
J.~Magueijo,
[arXiv:2012.05847 [gr-qc]].



\bibitem{Chopin}
L.~Smolin and C.~Soo,
Nucl. Phys. B \textbf{449}, 289-316 (1995). 


\bibitem{future}
J. Magueijo et al, ``The phenomenology of multi-time'', in preparation. 


\bibitem{LeeGL}
L.~Smolin,
Class. Quant. Grav. \textbf{33}, no.2, 025011 (2016).

\bibitem{eigenconstant}
J.~D.~Barrow and J.~Magueijo,
Phys. Rev. D \textbf{99}, no.2, 023509 (2019).

\bibitem{ebb}
J.~Magueijo and L.~Smolin,
Universe \textbf{5}, 84 (2019).


\bibitem{vikman}
P.~Jirou\v{s}ek, K.~Shimada, A.~Vikman and M.~Yamaguchi,
JCAP \textbf{04}, 028 (2021).

\bibitem{selftaught}
S.~Alexander, W.~J.~Cunningham, J.~Lanier, L.~Smolin, S.~Stanojevic, M.~W.~Toomey and D.~Wecker,
``The Autodidactic Universe,''
[arXiv:2104.03902 [hep-th]].

\bibitem{marina}
M.~Cort\^es and L.~Smolin,
Phys. Rev. D \textbf{90}, no.8, 084007 (2014); Phys. Rev. D \textbf{90}, no.4, 044035 (2014). 

\bibitem{rovelli}C.~Rovelli and M.~Smerlak,
Class. Quant. Grav. \textbf{28}, 075007 (2011).


\end{thebibliography}
\end{document}